# Virtual Coordinate Backtracking for Void Traversal in Geographic Routing

Ke Liu and Nael Abu-Ghazaleh

Binghamton University, SUNY
{kliu,nael}@cs.binghamton.edu

**Abstract.** Geographical routing protocols have several desirable features for use in ad hoc and sensor networks but are susceptible to voids and localization errors. Virtual coordinate systems are an alternative solution to geographically based routing protocols that works by overlaying a coordinate system on the sensors relative to well chosen reference points. VC is resilient to localization errors; however, we show that it is vulnerable to different forms of the void problem and have no viable complementary approach to overcome them. Specifically, we show that there are instances when packets reach nodes with no viable next hop nodes in the forwarding set. In addition, it is possible for nodes with the same coordinates to arise at different points in the network in the presence of voids. This paper identifies and analyzes these problems. It also compares several existing routing protocols based on Virtual Coordinate systems. Finally, it presents a new routing algorithm that uses backtracking to overcome voids to achieve high connectivity in the greedy phase, higher overall path quality and more resilience to localization errors. We show these properties using extensive simulation analysis.

## 1 Introduction

Traditional ad hoc routing protocols (e.g., AODV [1]), are not a good fit for Wireless Sensor Networks (WSN) for the following reasons. WSNS often require data dissemination patterns that are not efficiently mapped to the unicast connections assumed by ad hoc protocols. Further, nodes need to maintain routing state specific to destinations of active routes; this state may become invalid due to changes that are not near to the node. Finally, because of the need to maintain non-local state, this approach requires that the nodes have globally unique identifiers. As a result, this approach is not ideal for WSNs which favor routing protocols that support data-centric operation (e.g., global identifiers not necessary), localized interactions (e.g., maintaining only local state) and supporting arbitrary data-driven dissemination with in-network processing.

In contrast to traditional ad hoc protocols, Geographical routing algorithms [2, 3, 5, 6], provides attractive properties for WSNs. In these algorithms, nodes exchange location information with their neighbors. Packets addressed to a destination must provide its location. At every intermediate hop, the subset of the neighbors that are closer to the destination is called the forwarding set. Routing simply forwards a packet to one of the nodes in the forwarding set. This process is repeated greedily until the packet reaches the destination. Thus, interactions are localized to location exchange with direct neighbors and there is no need for global identifiers.

Geographical routing protocols suffer from significant problems under realistic operation. First, voids – intermediate nodes whose forwarding set (FS) relative to a destination is empty – can cause the greedy algorithm to fail [2–5]. Voids require a somewhat complex and inefficient complementary perimeter routing algorithm that is invoked when they are encountered [6]. Moreover, geographic routing has been shown to be sensitive to localization errors [7], especially in the perimeter routing phase [6, 8]; such errors can cause routing anomalies ranging from suboptimal paths to loops and failure to deliver packets. Making geographic routing practical is difficult [6].

Routing based on Virtual Coordinate Systems (VCS) has been recently proposed [13–16, 18]. A VCS overlays virtual coordinates on the nodes in the network based on their distance (typically in number of hops) from fixed reference points; the coordinates are computed via an initialization phase. These coordinates serve in place of the geographic location for purposes of Geographic forwarding; that is, in these algorithms the forwarding set is the set of nodes that are closer (based on different measures of coordinate distances) to the destination than the current node. Because it does not require precise location information, VCS is not sensitive to localization errors.

Further, it is argued that VCS is not susceptible to conventional voids because the coordinates are based on connectivity and not physical distance [14]. On the negative side, VCS may be sensitive to collisions and or signal fading effects in the initialization phase. Furthermore, the initialization phase requires a flood from each reference point.

The first contribution of the paper is to identify three problems that arise in VCS routing. In practice, the three problems occur in fairly common situations resulting in VCS failing to deliver packets. In the first problem, a set of neighbors are all of equal distance to the destination and greedy forwarding fails. In the second problem, we show that nodes with identical coordinates that are far from each other may arise in the presence of voids. In the third problem, a node is closer than any of its neighbors to a given destination, no matter how the distance is measured. We analyze the frequency of these problems using a number of random deployment scenarios of different densities.

The second contribution of the paper is to propose a hybrid routing protocol. The protocol uses Greedy Forwarding (GF) [2, 3] because of its superior path quality in the greedy mode. When voids are encountered, it switches to a VC based backup algorithm since, intuitively, VC is more effective in handling physical voids. However, since VC is susceptible to its own anomalies related to virtual coordinates, routing may still fail; we use a simple backtracking algorithm where the packet is forwarded backwards towards an anchor to address such possibilities. We show experimentally that it attains high delivery ratio and tolerates localization errors better than geographically based protocols. Our experiment results also show that for all scenarios we used, the Hybrid geographic and virtual coordinate Greedy Routing (HGR) provides a 100% reachability, and much better path quality than Greedy Perimeter Stateless Routing (GPSR) [2, 3]. The greedy nature of HGR makes its implementation practical and efficient.

The remainder of this paper is organized as follows. Section 2 provides an overview of the related works. After analyzing the problems of existing protocols and VCS in section 3, we present the design of HGR in section 4. In, Section 5 we present an experimental study analyzing the different routing protocols. Finally, Section 6 presents some concluding remarks.

## 2 Background and Related Work

Shortest Path (SP) routing is a commonly-used for sensor networks. In this protocol, data sinks send periodic advertisements that flood the network. As nodes receive the beacon, they set their next hop to the neighbor advertising the shortest distance to the sink. SP can provide the optimal path in terms of path length. However, it is a stateful (not stateless) and reactive protocol: for each destination, the forwarding path is needed before data transmission can begin. The storage it requires increases with the number of destinations in the network. Furthermore, it is vulnerable to mobility or other changes in the topology.

In ad hoc and sensor networks, stateless routing, where the routing state is independent of the traffic, is desirable. Geographical routing protocols with the stateless property [2, 3]; in the base mode, they use GF, where each node forwards packets to a neighbor that will bring the packet closest to the destination. Each node tracks only the location information of its neighbors. The *forwarding set* for a given destination is the subset of neighbors closer to the destination. GF proceeds by picking a node from this set, typically the closest one to the destination. If the forwarding set is empty, a *void* is encountered. Typically, a complementary algorithm is used to traverse the void. Face routing (or perimeter routing) is an approach based planar graph theory often used for void traversal. The general idea in face routing is to attempt to route around the void using a right hand rule that selects node around the perimeter of the void (details may be found in the original paper [2, 3]). Face routing stops when a node closer to the destination than the void origin is encountered; at this stage, operation switches back to greedy forwarding.

Since GPS devices are costly, they may not be feasible for sensor networks; often, localization algorithms are employed that significantly increase the uncertainty in the location estimate (e.g., [21, 23]). Both GF and face routing are susceptible to localization errors [7, 8]. While some approaches to tolerate location errors have been suggested, in general, this remains a weakness of this class of protocols. Further, the paths constructed by face routing are typically not the best path available to cross the void. Thus, additional routing protocols have attempted to optimize the face routing phase of operation [5, 4, 19].

Kim et al [6] recognize effects that arise in practice during geographic routing and suggest a protocol which uses more control packets to planarize the network. The algorithm requires more resources and is stateful.

Routing based on a coordinate system was first proposed by Rao et al [10]. Their algorithm requires a large number of nodes to serve as virtual coordinate anchors (sufficient to form a bounding polygon around the remaining sensors). A large number of anchor nodes increases the overhead as well as the state maintained by each node. The location estimated by the virtual coordinates is used for geographic routing. Their approach is more accurately described as a localization mechanism for use in an otherwise geographically based algorithm. It is unclear if face routing will be effective with a coarse-grained location estimate.

Caruso et al recently proposed the Virtual Coordinate assignment protocol (VCap) [16]; several similar protocols were proposed by others [13, 15, 14, 17, 18]. In this approach, coordinates are constructed in an initialization phase relative to a number of reference points. Following this initialization phase, packets can be routed using the Greedy Forwarding principles, replacing node location with its coordinates. The paper uses 3 reference points to assign the virtual coordinates, constructing a 3-dimensional VCS. The authors identified the *VC Zone* (several nodes assigned to the same virtual coordinates) problem and provided bounds on its size. (The original research work on VC Zone can be found in [9, 17] with more details.) The authors do offer a heuristic for situations where voids are encountered: they suggested a *local detour* to forward packet to some neighbor farther away to destination several times in hope of reaching some node with a different path to the destination; this approach may lead to longer paths as packets are misrouted. The evaluation in the paper shows that VCap performs worse than GPSR both in delivery ratio and path stretch.

Qing et al proposed a similar protocol to VCap with 4 reference nodes (4D) each located at a corner for a rectangle area [14]. The authors suggested a backtracking approach to packet delivery when facing any routing anomalies, which requires each hop in the forwarding path of each packet to be recorded. The authors did not analyze either why or when these anomalies happen. Although this backtracking approach converges, in the worst case, it will go through all the nodes in the network.

Using Manhattan-style distance (MD) in place of Euclidean distance (ED) was proposed by Rodrigo et al in BVR[18]. On a VCS with a very high number of reference nodes (typically 10 to 80), BVR suggested a different backtracking approach to send packets to the reference node closest to the destination when greedy forwarding fails. As we show in this paper, neither Manhattan distance or the one proposed in BVR[18] (we called semi-Manhattan distance) are a good measure of distance compared to a Euclidean distance. Having that many reference nodes require much more resources and may also hurt the performance of any practical sensor networks.

Papadimitriou and Ratajczak [12] conjecture that every planar 3-connected graph can be embedded on the plane so that greedy routing works. If this conjecture holds, then for planarized networks, a guaranteed greedy routing may exist. GEM [11] proposed the routing based on a virtual coordinate system. A virtual polar coordinate space (VPCS) is used for localizing each node in network. A tree-style overlay is then used for routing. Using the tree overlay results in poor path quality. Since it uses the VPCS to localize the network first, it tolerates only up to $10\%$ localization error [11].

## 3  Greedy Forwarding in VCS

Our approach is essentially Geographic Forwarding, with the use of VC when voids are encountered. This approach is motivated by the shortcomings in the planarization procedure of GPSR which has high cost and complexity, is susceptible to localization errors, and results in suboptimal routes. More specifically, perimeter routing requires calculation involving at least 2 hop information, and even 3 hops in some cases. Further, the calculation cost is high in terms of communication energy. Since GPSR requires the geographical information to make routing decisions, the accuracy of location information is the crucial factor. Although the GF works well under an up to $40\%$ of the localization error [7], face routing may fail with some very small localization error [8].

In contrast, Virtual coordinate systems are attractive because they are more resilient than geographic routing to localization errors, and because it is thought that they reduce the effect of voids. We use as the basis of description the Virtual Coordinate System (VCS) for introduced by VCap [16]; however, the problems identified and the solutions generalize to other virtual coordinate systems. The use of virtual coordinates introduces a different set of problems, which we identify in this section.

A network using SP with $N$ sinks can be considered an $N$-dimensional VCS as the distance to each of the sinks is tracked.

### 3.1 Number of Anchors

The authors [16] argue that for a 2 dimensional geographical coordinate system (GeoCS), a 3-dimensional VCS is sufficient to accomplish effective Greedy Forwarding (GF). We show that in practice this does not occur and VCS is susceptible to routing problems resulting in suboptimal paths, packet misrouting, or routing failure. These problems do not necessarily coincide with geographic voids: for example, we show that GF based on VCS may fail in a network without geographic voids.

Figure 1(a) shows a VCS for a network where 25 nodes are deployed along the vertices of a grid. The radio range makes each internal node have 9 one-hop neighbors. For example, node 13 has neighbors 7, 8, 9, 12, 14, 17, 18 and 19. Perimeter node 1 has 3 neighbors: 2, 6 and 7. The numbers at the left of each node are IDs, and the triples in brackets under each node represent their coordinate values. It can be show that there is no geographical void in this network (the forwarding set is never empty). VCS anchors are chosen consistent with the requirement described by its designers [16].

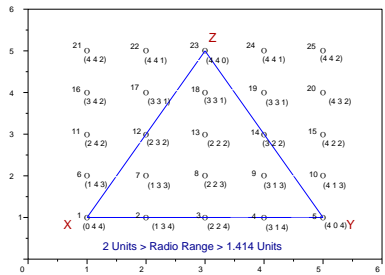 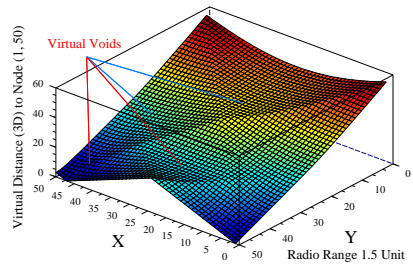

(a) 3-D VC for a Grid Deployment    (b) Disconnected VC Zone

**Fig. 1.** Routing Anomalies caused by VC Zone (3D VCS)

### 3.2 Expanded VC Zone Problem

Consider a packet *P* at node 11 destined to node 15. Table 1 shows the distances of each neighbors to the destination node 15. *P* is first forwarded to node 17 greedily. However, among the one-hop neighbors of 17 (node 11, 12, 13, 16, 18, 21, 22, 23), there are none with a shorter distance to node 15 measured in virtual coordinates – GF fails.

**Table 1.** The Euclidean distances of node 11, 17 and their neighbors on VCS to node 15

| node | 6 | 7 | 11 | 12 | 13 | 16 | 17 | 18 | 21 | 22 | 23 |
|------|---|---|----|----|----|----|----|----|----|----|----|
| distance | $\sqrt{14}$ | $\sqrt{11}$ | $2\sqrt{2}$ | $\sqrt{5}$ | 2 | $\sqrt{5}$ | $\sqrt{3}$ | $\sqrt{3}$ | 2 | $\sqrt{5}$ | $2\sqrt{2}$ |

Nodes with the same virtual coordinate value are called a *VC Zone*. This problem occurs when VC Zones cross a contour line (the lines connecting of the VCS anchors). Around these contour lines, the possibility of VC Zones that are larger than 2-hop across arises. The routing algorithm cannot deliver packets through the VC zone: we call this problem the *Expanded VC Zone* problem. One possible solution is to broadcast a path request within the VC zone.

### 3.3 Disconnected VC Zone Problem

The second problem occurs because it is possible for nodes with equal coordinates to occur in geographically disparate locations. Consider node 21 and 25; they have the same virtual coordinates but are not connected by any other nodes with the same virtual coordinates. We call this the *Disconnect VC Zone* problem. Node 22 and 24 are in the same situation. These nodes occur symmetrically around the contour line in a uniform deployment such as the grid. Note that the Expanded Zone problem is an instance of the Disconnected Zone problem where the two disconnected zones are neighbors.

If a data packet produced by node 21 needs to be delivered to node 25, GF will obviously fail. Further, data packets produced by any one-hop neighbors of node 21 can not be routed to node 25 either. Broadcasting within a VC zone cannot solve this problem because the VC zone is not connected. Even if the Z anchor goes to infinity (is arbitrarily far), a disconnection remains whose size is larger than the limit argued in the original VCap paper (2.3 times the radio range) [16]. Note that the limit in the original paper is derived under infinite density assumptions. We visualize this problem in figure 1(b), where 2500 nodes are deployed in $50 \times 50$ grids, each per one. 3 reference nodes are located at grid (1,1), (50, 1), (25, 50). For each point on the surface in figure 1(b), $x$ and $y$ values denote their physical location, $z$ values denote the Euclidean distances on VCS of each node to the node located in grid (1,50). The virtual distances to node(1,50) of node (50, 50) is 0, which leads to a virtual void caused by disconnected VC Zone. And there are still other virtual voids caused by expanded VC Zone.

We argue instead that the contour lines connecting VC anchors should be a polygon containing all nodes of the network inside it. If this occurs, then nodes occur only on one side of the contour line making both problems above impossible. Thus, a 4-dimensional

VCS is needed to guarantee the success of the greedy algorithm (with 4 corner nodes as anchors).

All the problems in VCS arise due to the following property: neighbors in VCS may not be neighbors geographically. In the presence of voids, no matter how carefully we choose the VCS anchors, this problem cannot be eliminated completely.

### 3.4 Effect of Distance Metric

Instead of using Euclidean distance to measure the distance between nodes in VCS, BVR [18] proposed a variant of Manhattan distance. We compare the effect of the different distance metrics using an example (Figure 2(a)). Node 1, 5, 12, 16 are 4 reference nodes. Suppose the data source is node 14 in this figure, while destination is node 3. All the distances of any type of the neighbors of node 14 is far away from the destination than itself, shown in table 2. We also visualize this problem in figure 2(b) with similar scenario as figure 1(b) except a physical void in the center. As we can see, the Manhattan-style distance does not help the distance measurement, while it may make the problem much worse. Euclidean distance performs better than the Manhattan-styled distance but still with some virtual voids (figure 3).

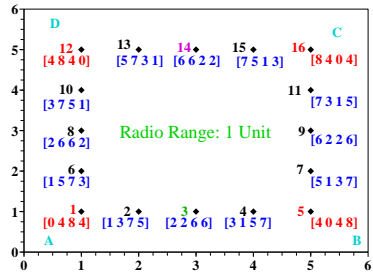 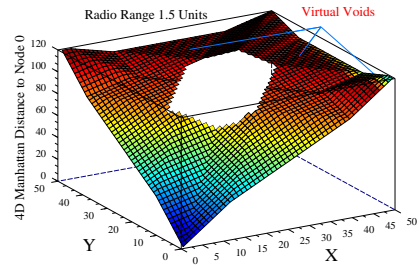

(a) 4D VCS with physical void  (b) Virtual Voids with Manhattan Distance

**Fig. 2.** Anomaly despite of Diversity of Distances (4D VCS)

## 4 Hybrid Greedy Routing (HGR)

Our simulation result show that for any path GF constructs successfully, the path quality is almost identical to the optimal found by SP; moreover, this phase of the algorithm is tolerant to localization errors. Thus, our algorithm uses Geographic Forwarding as the base. HGR uses virtual coordinates for void avoidance because they naturally protect against geographical voids. However, in the presence of the problems outlined in the previous section, VC routing may fail. As a result, we use a simple backtracking technique in response to VC routing anomalies.

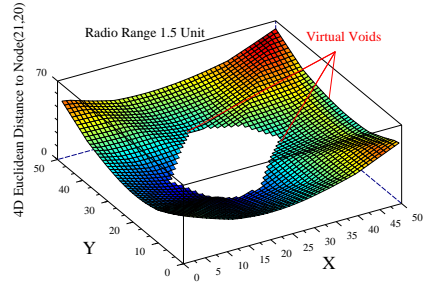

**Fig. 3.** Virtual Voids measured in Euclidean Distance (4D VCS)

**Table 2.** Different types of Distances to Node 3

| Distance Diversity | Node 12 | Node 13 | Node 14 | Node 15 | Node 16 |
|---|---|---|---|---|---|
| Euclidean (VCap, LCR) | $\sqrt{80}$ | $\sqrt{68}$ | $\sqrt{50}$ | $\sqrt{68}$ | $\sqrt{80}$ |
| Manhattan | 16 | 16 | 16 | 16 | 16 |
| Semi-Manhattan (BVR) $\delta+$ | 8 | 8 | 8 | 8 | 8 |
| Semi-Manhattan (BVR) $\delta-$ | 8 | 8 | 8 | 8 | 8 |

### 4.1 Void Avoidance Phase

To resolve the void problem, HGR uses virtual coordinates. The VCS of the network is initialized by the same procedure introduced by VCap [16]. The intuition of HGR is that if any node in the network with a VC value smaller than infinity, is reachable. Once a local maximum (void) is encountered, HGR switches to *hybrid-mode*. In the hybrid mode, it picks one of the coordinate axes and attempts to route using it towards the destination. If a point is reached where no neighbor closer to the destination is found, then backtracking is necessary.

We elect to carry out VC as well as backtracking in an axis by axis basis – alternative approaches are possible and will be a topic of future research. More specifically, if we reach a point along the current axis where no node leads closer to the destination on the same axis, the direction is reversed and we backtrack. In the current implementation, we backtrack until a void (relative to the same axis) is found while backtracking, or we reach coordinate 0. Alternative approaches for terminating backtracking are also possible. If either of the cases occur, we switch to the next axis and repeat the process. If all three axes are exhausted, the routing fails[1]. At any point in the algorithm if a node geographically closer to the destination than the node at which the hybrid mode

---

[1] A backup algorithm such as localized flooding may be used here, but in practice, we observe very few failures

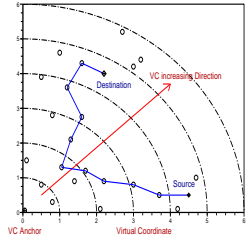

**Fig. 4.** HGR routing Demo

is entered is found, the protocol switches back to the greedy mode (the void had been traversed).

The hybrid mode operates as follows. When a packet is faced by a void, it is labeled as hybrid-mode and the geographical coordinates of the current node, $GC_{initial}$, are entered into the packet header. The node also records the dimension index $i$ (initially, $i = 1$) and the VC direction for dimension $i$ into the header, where the VC direction is decided by the VC of the current node and the destination ($VC^i_{initial}$ and $VC^i_{sink}$) according to the function:

$$direction^i = -1 \; if \; VC^i_{sink} \leq VC^i_{entering} \; else \; 1$$

On receiving a *hybrid-mode* data packet, (ED stands for Euclidean distance)

1. **If** $ED_{entering,sink} > ED_{current,sink}$, to label *greedy-mode* to the data packet, switch back to using greedy forwarding, **else**:
2. **If** sink is in the neighbor list of current node, forward the data packet to sink and routing succeeds, **else**:
3. **If** $VC^i_{last_hop} = VC^i_{current}$, goto 4), **else**: Among all neighbors $n$ with $VC^i_n = VC^i_{current} - direction^i$ (same $VC^i$ as last hop), **if** $ED_{N,sink} = min(ED_{n,sink})$ and node $N$ is not the last hop, then $direction^i = -direction^i$, forward to $N$; **otherwise**:
4. Among all neighbors $n$ with $VC^i_n = VC^i_{current}$ (same $VC^i$ as current node), **if** $ED_{N,sink} = min(ED_{n,sink})$ and $N \neq current$, forward to $N$; **otherwise**:
5. Among all neighbors $n$ with $VC^i_{current} + direction^i$, forward the data packet to the Node $N$ where $ED_{N,sink} = min(ED_{n,sink})$ for all $n$, **otherwise**: **if** no such neighbor exists, reverse the direction: $direction^i = -direction^i$, goto 2); **If** $direction^i$ has been reversed once in previous routing procedure (either current node, or other nodes, but a reverse in Step 3 does not count), increase the coordinate index $i$, if $i \leq max(dimension)$ goto 2); **otherwise**, label this data packet with *HGR fails*, **Quit**;

Note that if current node and destination have finite virtual coordinate values, they should be connected in the network. Figure 4 shows a sample path obtained by HGR.

## 5 Experiment

In this section, we present an experimental evaluation that illustrates the problems with existing geographical and VC protocols. The evaluation also characterizes the performance of the proposed HGR protocol.

To allow scalability to very large networks, we use a custom simulator written for this study; the simulator abstracts away the details of the channel and the networking protocols which may affect performance such as the reach-abilities of routing protocols. Our results validate successfully with the NS-2 simulator.

We study both random and controlled deployment. Each point represents the average of 30 scenarios of 200 nodes that are deployed in a $1000 \times 1000 m^2$ area; the number of scenarios was sufficient to tightly bound the confidence intervals (for random scenarios, each node's location was generated uniformly in the whole simulation area). We simulate the different densities by varying the radio transmission range. For every scenario, reachability is determined by testing whether a packet can be delivered between each pair of nodes in the network. Recall that the stateful SP is the optimal routing in terms of number of hops; for this reason it is used to derive the ideal performance in terms of path quality.

We also implemented the GPSR (with GG and RNG planarization algorithms)[3, 6], Shortest Path (SP), Greedy Forwarding on VCS [16, 14] and BVR on 4D VCS [18], and study their performance against HGR. To enable a fair comparison, we use the same number of anchors for all protocols (which, in fairness to BVR, is much lower than the number of beacons the authors recommend). However, we believe a small number of anchors is essential for deeply embedded sensor networks.

### 5.1 Greedy Forwarding (GF)

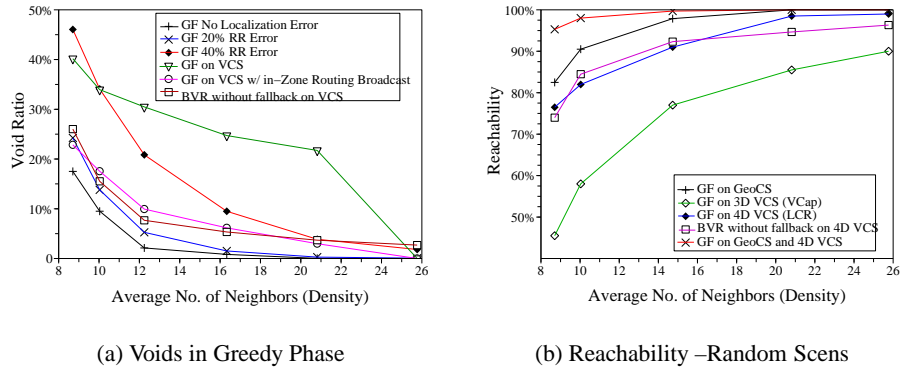

(a) Voids in Greedy Phase

(b) Reachability –Random Scens

**Fig. 5.** Greedy Forwarding Analysis

**Analysis of Void Frequency** This first study shows the frequency of occurrence of voids for both geographic and VC routing (based either virtual Euclidean distance or semi-Manhanttan distance). Previously, it was demonstrated that voids occur in different situations for the two types of protocols. Figure 5(a) shows the ratio of node pairs facing void problem. In general, we can see that sparse networks suffer from void problem much more than dense ones. A primary observation from this graph is that the frequency of voids is significantly higher in VCS systems than in Geographic Coordinate (GeoC) ones even when using 4 coordinate axes. This results shows that while the analysis in the original VCap [16] paper may apply under asymptotic conditions (infinite density), it does not hold under practical situations. It is clear that VCap on its own does not improve performance relative to pure GF even with 4 coordinates such as LCR without backtracking [14].

Localization errors affect the performance of GF on GeoCS significantly. Since the VCS does not need the location information to initialize, it does not suffer from this problem. However, the expanded VC zone plays an important effect even when network density is high. This can be seen in the graph where the zone broadcast is implemented (figure 5(a), curve labelled as "GF on VCS w/ in-Zone Routing Broadcast"); zone broadcast floods a packet in a VC zone, which eliminates the expanded zone problem but incurs some additional overhead. This curve suffers much fewer voids than just regular VC. The remaining voids in this curve are due to the disconnected VC Zone problem.

**Analysis of Greedy Forwarding** Figure 5(b) shows the reachability of all pairs in the same random deployment scenario, using only GF as routing (void traversal is not compared here). GF based on a 3D VCS (VCap) shows the worst reachability. The 4D VCS (LCR) shows a much higher reachability than 3D VCS, but still worse than GF on GeoCS. We also use a combination of the GeoCS and 4D VCS: GF on GeoCS first, when it fails, GF on 4D VCS is used. The result shows this combination works much better than any one independently. The reachability of it is higher than $95\%$ even in a sparse network which leads much smaller cost of backtracking. Although our experiment results show a higher-dimensional VCS working better than a lower-dimensional one for GF routing, we found that this does not hold beyond 4 dimensions (graph not shown due to space limitation).

### 5.2 HGR performance

**Randomly Deployed Networks** Figure 6(a) shows the path quality obtained across all nodes under different densities. SP routing provides the optimal solution, which cannot be obtained by any stateless greedy solution in general. The performance of HGR is much better than GPSR with either GG or RNG planarization. When the densities of networks go higher, less voids happen, leading to similar overall path qualities. The average path length of BVR is the highest. The reason may be that in BVR the backtracking path is much longer since it needs to forward packet to some one of the reference node.

Figure 6(b) shows the quality of only the paths facing void problems. As the density goes higher, loss of efficiency results due to the planarization algorithm (which forces using the nearest neighbor). In contrast, HGR operates greedily even in the void traversal/hybrid mode. As a result, HGR performs well while the average performance of GPSR suffers. Figure 7(a) shows a sample path between 2 nodes in one of the 30 networks, generated by different routing protocols with radio range as $150m$.

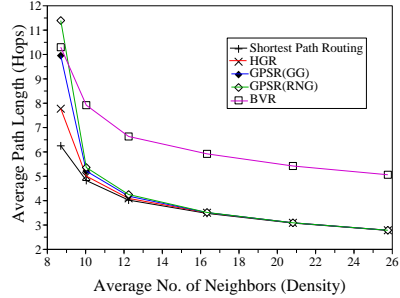
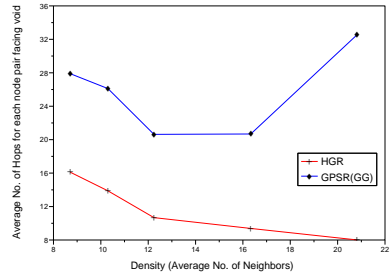

(a) Overall Path Quality  (b) Path Quality of Void Traversal

**Fig. 6.** Path Stretch

**Performance under Frequent Voids** To study the behavior of the protocols under voids, we also create scenarios where 150 nodes are randomly deployed in a "C" region around the border of the area. In this case, a large portion of the paths are faced by voids.

Figure 7(b) shows one path in a "C" model network with radio range as $200m$. The path length of HGR is longer than that of GPSR in Euclidean distance, but much shorter in number of hops.

Figure 8 shows the average path length of routing protocols in the 20 randomly deployed "C" networks. Once the radio range is too small to cross the void, greedy forwarding faces voids. HGR performs much better than any of the GPSR flavors and BVR, roughly approximating the optimal solution. BVR outperforms GPSR as well; virtual coordinate routing is effective in traversing physical voids; this scenario does not result in creating many VC voids.

**Impact of Localization error** We also study the impact of localization error on the different protocol. For routing, localization values uniformly distributed in a circle of radius $ratio \times range$ around the correct location are generated (resulting in average error of $ratio \times range$. GPSR may fail when the localization error is big, in either the greedy phase (causing an unnecessary switch to face routing), or in the face routing phase (causing routing failure). HGR is also susceptible to routing anomalies when it uses the geographic location in the greedy phase in in certain instances in the hybrid phase. We observed that the frequency of routing errors is much higher in GPSR compared to HGR, which tolerates errors well in all but very sparse scenarios. In order to study the effect on path quality, we planarized the graph based on the symmetric connection of neighbors, used in [6], that is not affected by localization errors (which benefits only GPSR), and study the impact of localization error here. HGR is still carried distributedly as before. Figure 9 shows the average path length in the 30 randomly deployed networks, with error of $ratio = 20\%$ radio range (effective location is 20%

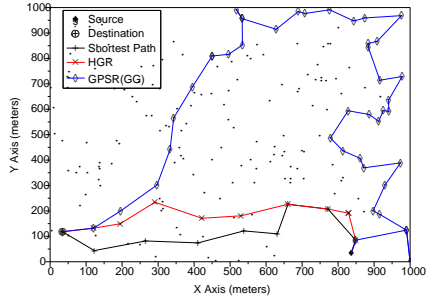
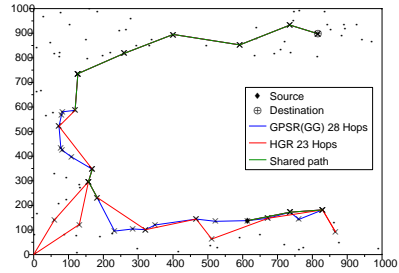

(a) Sample Paths by Different Protocols (RR 150m)

(b) Paths in "C" Networks (RR 200m)

**Fig. 7.** Sample HGR Path vs GPSR Path Anomalies

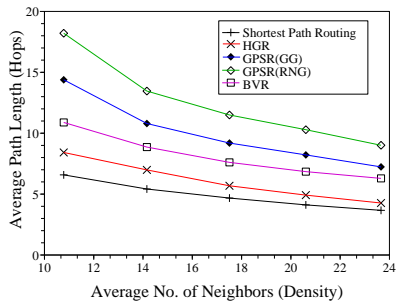
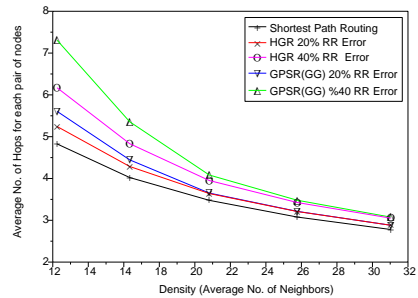

**Fig. 8.** Path Quality in "C" Networks

**Fig. 9.** Impact of Localization Errors

radio range away from the actual location) and $40\%$ radio range under different densities. The reachability of HGR is still 100% and the path quality is much better than the centrally planarized GPSR.

## 6 Conclusion

In this paper, we first demonstrate problems that arise with Virtual Coordinate based routing and show that contrary to published conclusions, it is quite vulnerable to voids that arise during greedy operation. More specifically, we identify the expanded VC Zone problem, where nodes of equal VC coordinates span a multiple hop zone. We also identify the disconnected VC zone problem where nodes sharing the same VC value occur in geographically disparate locations. We show experimentally that these problems have a considerable effect on reachability provided by VCS algorithms.

The second contribution of the paper is to present a hybrid routing protocol that replaces the traditional face routing phase used for void traversal in geographic routing protocols with one based on virtual coordinates. We use a simple dimension by dimension hueristic with backracking along the same dimension as an example of this type of approach. We show that the resulting algorithm significantly outperforms GPSR and BVR in terms of reachability and path quality. It is also significantly more resilient to localization errors, especially those that affect the perimeter routing phase of the algorithm. Nevertheless, we believe that improved hueristics in the void traversal phase are still possible. This is a topic of our future research.

## References

1. C. E. Perkins and E. M. Royer, *Ad hoc On-Demand Distance Vector Routing*, in Proceedings of the 2nd IEEE Workshop on Mobile Computing Systems and Applications, February 1999
2. P. Bose, P. Morin, I. Stojmenovic and J. Urrutia, *Routing with guaranteed delivery in ad hoc wireless networks*, DIAL M99, Auguest 1999
3. B. Karp and H.T. Kung, *GPSR: Greedy Perimeter Stateless Routing for Wireless Networks*, MobiCom 2000
4. Q. Fang, J. Gao and L. Guibas, *Locating and Bypassing Routing Holes in Sensor Networks*, INFOCOM 2004
5. Sophia Fotopoulou-Prigipa and A. Bruce McDonald, *GCRP: Geographic Virtual Circuit Routing Protocol for Ad Hoc Networks*, MASS 2004
6. Young-Jin Kim, Ramesh Govindan, Brad Karp and Scott Shenker, *Geographic Routing Made Practical*, the Second USENIX/ACM Symposium on Networked System Design and Implementation(NSDI'05), May 2005
7. T. He, C. Huang, B. Blum, J. A. Stankovic and T. Abdelzaher, *Range-Free Localization Schemes for Large Scale Sensor Networks*, MobiCom 2003
8. K. Seada, A. Helmy and R. Govindan, *On the Effect of Localization Errors on Geographic Face Routing in Sensor Networks*, IPSN 2004
9. Samir Khuller, Balaji Raghavachari and Azriel Rosenfeld, *Landmarks in graphs*, Journal of Discrete Appl. Math., volume 70, page 217–229, 1996
10. A. Rao, S. Ratnasamy, C. Papadimitriou, S. Shenker and Ion Stoica, *Geographic Routing without Location Information*, MobiCom 2003
11. J. Newsome and D. Song, *GEM: Graph EMbedding for Routing and Data-Centric Storage in Sensor Networks Without Geographic Information*, SenSys'03, Nov. 2003
12. C. H. Papadimitriou and D. Ratajczak, *On a Conjecture Related to Geometric Routing*, ALGOSENSORS 2004
13. T. Moscibroda, R. O'Dell, M. Wattenhofer, R. Wattenhofer, *Virtual Coordinates for Ad hoc and Sensor Networks*, ACM Joint Workshop on Foundations of Mobile Computing (DIALM-POMC), October 2004
14. Qing Cao and Tarek F. Abdelzaher, *A Scalable Logical Coordinates Framework for Routing in Wireless Sensor Networks*, RTSS 2004
15. D. M. Nicol, M. E. Goldsby and M. M. Johnson, *Simulation Analysis of Virtual Geographic Routing*, in Proceedings of the 2004 Winter Simulation Conference
16. A. Caruso, S. Chessa, S. De and A.o Urpi, *GPS Free Coordinate Assignment and Routing in Wireless Sensor Networks*, INFOCOM 2005
17. Mirjam Wattenhofer, Roger Wattenhofer and Peter Widmayer, *Geometric Routing Without Geometry.*, SIROCCO 2005
18. R. Fonseca, S. Ratnasamy, J. Zhao, C. T. Ee, D. Culler, S. Shenker and I. Stoica, *Beacon Vector Routing: Scalable Point-to-Point Routing in Wireless Sensornets*, NSDI'05, 2005

19. Q. Fang, J. Gao, L. J. Guibas, V. de Silva and L. Zhang, *GLIDER: Gradient Landmark-Based Distributed Routing for Sensor Networks*, INFOCOM 2005
20. R. Nagpal, H. Shrobe and J. Bachrach, *Organizing a global coordinate system from local information on an ad hoc sensor networks*, IPSN 2003
21. D. Niculescu and B. Nath, *Ad Hoc Positioning System (APS)*, GlobalCom 2001
22. J. Hightower and G. Borriella, *Location Systems for Ubiquitous Computing*, IEEE Computer, V34, 57-66, 2001
23. A. Haeberlen, E. Flannery, A. M. Ladd, A. Rudys, D. S. Wallach and L. E. Kavraki, *Practical Robust Localization over Large-Scale 802.11 Wireless Networks*, MobiCom 2002